\documentclass{article}
\usepackage{graphicx}
\usepackage{amsmath}
\usepackage{amsfonts}
\usepackage{amssymb}

\begin{document}

\title{Non-Holonomic Control I}
\author{E. Brion\\\emph{Laboratoire Aim\'{e} Cotton, }\\\emph{CNRS II, B\^{a}timent 505, }\\\emph{91405 Orsay Cedex, France.}
\and V.M. Akulin\\\emph{Laboratoire Aim\'{e} Cotton, }\\\emph{CNRS II, B\^{a}timent 505, }\\\emph{91405 Orsay Cedex, France.}
\and D. Comparat\\\emph{Laboratoire Aim\'{e} Cotton, }\\\emph{CNRS II, B\^{a}timent 505, }\\\emph{91405 Orsay Cedex, France.}
\and I. Dumer\\\emph{College of Engineering, }\\\emph{University of California, }\\\emph{Riverside, CA 92521, USA. }
\and V. Gershkovich\\\emph{Institut des Hautes Etudes Scientifiques,}\\\emph{ Bures-sur-Yvette, France. }
\and G. Harel\\\emph{Department of Computing, }\\\emph{University of Bradford, }\\\emph{Bradford, West Yorkshire BD7 1DP, United Kingdom. }
\and G. Kurizki\\\emph{Department of Chemical Physics, }\\\emph{Weizmann Institute of Science, }\\\emph{76100 Rehovot, Israel. }
\and I. Mazets\\\emph{Department of Chemical Physics, }\\\emph{Weizmann Institute of Science, }\\\emph{76100 Rehovot, Israel. }\\\emph{A.F. Ioffe Physico-Technical Institute, }\\\emph{194021 St. Petersburg, Russia. }
\and P. Pillet\\\emph{Laboratoire Aim\'{e} Cotton, }\\\emph{CNRS II, B\^{a}timent 505, }\\\emph{91405 Orsay Cedex, France.}}
\maketitle
\begin{abstract}
In this paper, we present a universal control technique, the non-holonomic control, which allows us to impose any arbitrarily prescribed unitary evolution to any quantum system through the alternate application of two well-chosen perturbations. 
\end{abstract}

\section{Introduction}

Quantum control  is very topical and interests many different fields of
contemporary physics and chemistry, such as Molecular Dynamics in laser fields
and Quantum Optics \cite{Tannor:85, PK02, Shapiro:86, Peirce:88, Law:96}. A few examples of control of the quantum state by conditional
measurements \cite{Vogel:93, Harel:96}, by adiabatic transport or by
unitary evolution have been already proposed for the particular quantum system
of atoms interacting with quantized electromagnetic field in a single-mode
resonator. In parallel, a theoretical framework of quantum control has been
built up : in particular, several classes of problems have been distinguished
such as the control of quantum evolution, quantum state, or density matrix ;
besides, the mathematical necessary conditions for their feasibility have been
derived in the context of the theory of Lie groups \cite{JS72, Huang:83}.
Moreover, theoretical methods, mainly based on optimization techniques, have been designed to exhibit the explicit values of the control parameters.

Most of these traditional approaches rely either on a known or intuitively guessed particular solution which can be further optimized with respect to a given cost functional, through variational schemes \cite{BS90}. By contrast, identifying the convergence domain of the standard algorithms through straightforward exploration of the entire multidimensional space of the control parameters implies numerical work, the complexity of which grows exponentially with the dimensionality of the system. Though, as we shall see in this paper, an original method, called non-holonomic bang-bang control \cite{Harel:99}, solves this problem for complex enough quantum systems so that no separable subsystems can be singled out : in terms of classical mechanics, such systems do not have holonomic constraints. The physical idea is to alternately apply two distinct perturbations $\widehat{P}_{a}$ and $\widehat{P}_{b}$ during pulses, the timings of which play the role of control parameters and are determined by solving the \textquotedblright inverse Floquet problem\textquotedblright. Equivalently, one can fix the duration of the pulses and choose the strength of the perturbations as the free parameters one has to adjust in order to achieve the control objective. Actually, the convergence of our algorithm results from an unsuspected simplification emerging from the Random Matrix Theory. Indeed, it relies on the algebraic properties of the $N^{th}$ roots of the identity matrix, the spectra of which resemble to those of random unitary matrices which obey the Dyson distribution law. 

This paper is organized as follows. In the first part, we recall the broad context of the theory of quantum control. In the second part, we present the non-holonomic control technique, and provide all the algorithmic tools which allow one to implement it. 

\section{Quantum Control}

Quantum control is a very topical issue in contemporary physics. The needs for
control are particularly sensitive in Molecular Dynamics and Quantum Optics,
and are quite various : indeed, one should not speak of quantum control, but
rather of quantum controls. Actually, four different types of problems have
been identified in the literature \cite{BS90, SPS03}: the control of pure state, the control of
density matrix, the control of observable and, finally, the control of the
evolution operator. For each of these problems, the goal is the same, that is
to impose the considered characteristics an arbitrarily chosen value.

To achieve a control objective, one has to perturb the system, since its
natural evolution usually results in too restrictive a dynamics. The control
Hamiltonian $\widehat{H}\left(  t\right)  $ comprises thus the unperturbed
Hamiltonian $\widehat{H}_{0}$ as well as $M$ Hamiltonians of the form
$C_{m}\left(  t\right)  \widehat{P}_{m}$, which can stand for the interaction
Hamiltonians of the system with $M$ classical fields, the amplitudes of which,
denoted by $C_{m}\left(  t\right)  $, are controlled by an external operator%
\[
\widehat{H}\left(  t\right)  =\widehat{H}_{0}+\sum C_{m}\left(  t\right)
\widehat{P}_{m}.
\]
The functions $\left\{  C_{m}\left(  t\right)  \right\}  $ play the role of
the control parameters one has to adjust in order to achieve the desired
control process. In other words, any problem of control can always be
translated into the following form : for the physical system considered,
perturbed through given fields of controllable amplitudes $\left\{
C_{m}\left(  t\right)  \right\}  $, one looks for the values of the control
paramaters $\left\{  C_{m}\left(  t\right)  \right\}  $ such that a specific
characterics (quantum state, density matrix, observable, evolution operator)
takes an arbitrarily prescribed value.

All the objectives are not feasible. Some of them are forbidden by the
unitarity of the evolution operator for closed systems : for example, one
cannot change the eigenvalues of the density matrix through a Hamiltonian
process of control. This kind of constraints is often referred to as
\emph{kinematical constraints} \cite{SSL02}. But there also exist \emph{dynamical
constraints} which stem from the algebraic properties of the Hamiltonians
$\left\{  \widehat{P}_{m}\right\}  $. Indeed, the evolution operator
\[
\widehat{U}\left(  t\right)  =\mathcal{T}\left\{  e^{-\frac{i}{\hbar}\int
_{0}^{t}\widehat{H}\left(  \tau\right)  d\tau}\right\}
\]
where $\mathcal{T}$ denotes the chronological product, belongs to the Lie
group obtained by exponentiation of the Lie algebra generated by the operators
$\left\{  i\widehat{H}_{0},i\widehat{P}_{m}\right\}  $. So, it appears clearly
that the feasibility of a particular problem of control in a specific physical
situation, defined by the Hamiltonians $\left\{  \widehat{P}_{m}\right\}  $, is
closely related to the properties of this algebra : for example, if
one wants to completely control the evolution operator of a quantum system,
that is if one wants to be able to give the operator $\widehat{U}$ any
prescribed value, one must perturb it in such a way that the operators
$\left\{  i\widehat{H}_{0},i\widehat{P}_{m}\right\}  $ generate the whole Lie
algebra $u\left(  N\right)  $ which provides, through exponentiation, the
whole Lie group $U\left(  N\right)  $ \cite{JS72, RSDRP95}. Necessary mathematical conditions also
exist for the other types of control problems and can be found in literature \cite{SPS03} : these
conditions are obviously weaker than the previous one, since the evolution
controllability automatically implies all the other ones.

The feasibility of a control problem can thus be decided through mathematical
criteria established in the context of the Lie group theory. But the explicit
values of the control parameters achieving the desired control objective still
remain to be found. In other words, once the existence of a solution has been
proved, one has to find it explicitly. Different methods have been proposed,
such as optimal control \cite{OKF98, SGL00, PK02}, which mainly employ optimization techniques : the
idea is to write down a functional the optimization of which gives the best
compromise between all the constraints. A purely algebraic approach \cite{SGRR01}, based on
the decomposition of the arbitrary desired evolution on the Lie group, is also
possible, but rapidly leads to intractable computations as the dimension of
the state space increases.

In the next section, we describe a universal control technique, called
non-holonomic control \cite{Harel:99}, which allows one to control the evolution operator of
an arbitrary quantum system through the alternate applications of two
well-chosen perturbations. Two equivalent sets of control parameters are
possible : one can choose to freeze the amplitudes of the perturbations and
let the pulse timings play the role of free parameters, or, on the contrary,
freeze the pulse timings and adjust the amplitudes on each control pulse. We
shall present the algorithm which allows us to compute the explicit values of
control parameters, and emphasize the algebraic reasons which assure its convergence.

\section{Control of the evolution through non-holonomic control}

Let us consider an $N$-dimensional quantum system of unperturbed Hamiltonian
$\widehat{H}_{0}$. Our goal is to control its evolution operator $\widehat{U}%
$, \emph{i.e.} to be able to achieve any arbitrary evolution $\widehat
{U}_{arbitrary}\in U\left(  N\right)  $.\
\begin{figure}
[ptb]
\begin{center}
\includegraphics[
height=1.9199in,
width=2.7717in
]%
{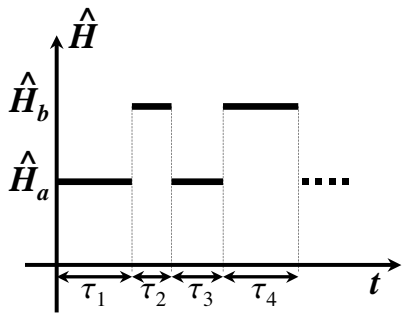}%
\caption{Pulsed shape of the control Hamiltonian.}%
\label{Fig1}%
\end{center}
\end{figure}
To this end, we alternately apply two physical perturbations, of Hamiltonians
$\widehat{P}_{a}$ and $\widehat{P}_{b}$, during $N^{2}$ pulses the timings of
which are denoted by $\left\{  \tau_{k}\equiv t_{k}-t_{k-1}\right\}
_{k=1,\ldots,N^{2}}$ ($t_{0}=0$ and $t_{N%
{{}^2}%
}=T$ correspond to the beginning and the end of the control sequence,
respectively) : the total Hamiltonian takes the following pulsed shape (cf
fig. \ref{Fig1})%
\[
\widehat{H}(t)=\widehat{H}_{0}+C_{a}\left(  t\right)  \widehat{P}_{a}%
+C_{b}\left(  t\right)  \widehat{P}_{b}%
\]
where%
\begin{align*}
C_{a}\left(  t\right)   &  =1\text{, }C_{b}\left(  t\right)  =0\text{\ and
}\widehat{H}(t)=\widehat{H}_{0}+\widehat{P}_{a}\equiv\widehat{H}_{a}\text{
\ for }t\in\left[  t_{2k},t_{2k+1}\right] \\
C_{a}\left(  t\right)   &  =0\text{, }C_{b}\left(  t\right)  =1\text{\ and
}\widehat{H}(t)=\widehat{H}_{0}+\widehat{P}_{b}\equiv\widehat{H}_{b}\text{
\ for }t\in\left[  t_{2k-1},t_{2k}\right] \\
\text{for \ }k  &  =1,\ldots,N^{2},
\end{align*}
and the total evolution operator is%
\[
\widehat{U}\left(  \left\{  \tau_{1},\ldots,\tau_{N^{2}-1},\tau_{N^{2}%
}\right\}  \right)  =e^{-\frac{i}{\hbar}\widehat{H}_{b}\tau_{N^{2}}}\cdot
e^{-\frac{i}{\hbar}\widehat{H}_{b}\tau_{N^{2}-1}}\ldots e^{-\frac{i}{\hbar
}\widehat{H}_{a}\tau_{1}}.
\]
where we have implicitly assumed that $N$ is even.

Our control problem can thus be translated into the following form : given
$\widehat{U}_{arbitrary}\in U(N)$, an arbitrary unitary operator, we want to
find a time vector $\overrightarrow{\tau}=\left(
\begin{array}
[c]{c}%
\tau_{1}\\
\vdots\\
\tau_{N^{2}-1}\\
\tau_{N^{2}}%
\end{array}
\right)  $ such that
\begin{equation}
\widehat{U}\left(  \overrightarrow{\tau}\right)  =\widehat{U}_{arbitrary}.
\label{pbcont}%
\end{equation}

As we said previously, for a solution to exist the operators $\left\{
i\widehat{H}_{a},i\widehat{H}_{b}\right\}  $ must generate the whole Lie
algebra $u\left(  N\right)  $ : this is called the ''bracket generation condition''. This property can be checked directly as long
as the dimension $N$ is not too big : one simply computes the commutators of
all orders of $i\widehat{H}_{a}$ and $i\widehat{H}_{b}$ and stops as soon as
they generate $u\left(  N\right)  $. But when $N$ becomes large, direct
computation is intractable. In that case, one can simply check the following
sufficient condition, suggested by V. Kac, according to which the system
becomes non-holonomic, that is completely controllable, when the
representative matrix of $\widehat{H}_{b}$ in the eigenbasis of $\widehat
{H}_{a}$ has no off-diagonal zeros.

Once the previous criterion is checked, one has to compute the time vector
$\overrightarrow{\tau}$ solution of Eq.(\ref{pbcont}). The method consists
first in determining the time vector $\overrightarrow{\tau}^{(0)}$ such that%
\[
\widehat{U}\left(  \overrightarrow{\tau}^{(0)}\right)  =\widehat{I},
\]
and then iteratively approaching the time vector $\overrightarrow{\tau}$
through a Newton-like technique.

\begin{figure}[ptbh]
\begin{center}
\includegraphics[
width=2.5 in,
]{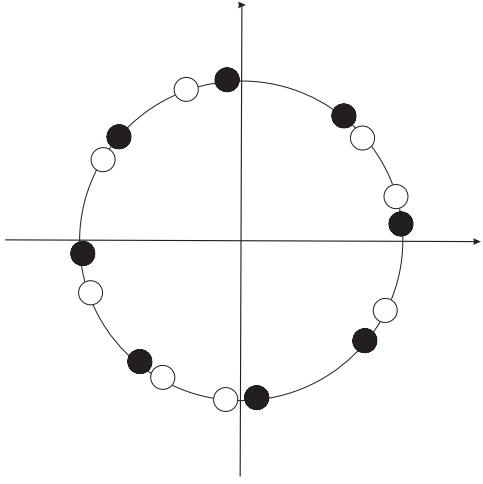}
\end{center}
\caption{Spectrum of a random unitary matrix (white circles) resulting from
the repulsion of the eigenvalues on a unit circle is shown vs the eigenvectors
of $N^{th}$ root of the identity matrix (black circles) multiplied by an
unimportant phase factor.}%
\label{Fig2}
\end{figure}

The straightforward way to compute $\overrightarrow{\tau}^{(0)}$\ would be to
minimize the functional%
\[
F\left(  \overrightarrow{\tau}\right)  =\Vert\widehat{U}\left(
\overrightarrow{\tau}\right)  -\widehat{Id}\Vert^{2}%
\]
with respect to $\overrightarrow{\tau}$. However, $F$ presents many local
minima which make its optimization uneasy. Nevertheless there exists an alternative
method based on the algebraic properties of the $N^{th}$ roots of the identity matrix. The idea is to look for $N$ parameters $\{T_{k}\}_{k=1...N}$ such that
\begin{equation}
e^{-\frac{i}{\hbar}\widehat{H}_{b}T_{N}}\cdot e^{-\frac{i}{\hbar}\widehat
{H}_{a}T_{N-1}}\ldots e^{-\frac{i}{\hbar}\widehat{H}_{a}T_{1}}=\widehat{I}%
^{\frac{1}{N}},
\end{equation}
where $\widehat{I}^{\frac{1}{N}}$ is an $N^{th}$ root of the identity matrix,
\emph{i.e.} a matrix the eigenvalues of which coincide with the $N^{th}$ roots
of the unity $\left\{  e^{ik\frac{2\pi}{N}}\right\}  _{k=0,\ldots,\left(
N-1\right)  }$ ; in other words, $\widehat{I}^{\frac{1}{N}}$ is a matrix of the
form
\[
\widehat{I}^{\frac{1}{N}}=\widehat{M}^{-1}\cdot\left[
\begin{array}
[c]{cccc}%
1 & 0 & \cdots & 0\\
0 & e^{i\frac{2\pi}{N}} & \cdots & 0\\
\vdots & \vdots & \ddots & \vdots\\
0 & 0 & \cdots & e^{i\left(  N-1\right)  \frac{2\pi}{N}}%
\end{array}
\right]  \cdot\widehat{M},
\]
where $\widehat{M}$ is a unitary matrix. To compute the $T_{k}$'s, we use the
following algebraic property : if $P_{\widehat{U}}(\lambda)\equiv\sum
_{j=0}^{N}a_{j}\lambda^{j}$ denotes the characteristic polynomial of a unitary
matrix $\widehat{U}$, then $\sum_{j=0}^{N}\left\vert a_{j}\right\vert ^{2}%
\geq2$ and the equality is achieved iff $\widehat{U}$\ is an $N^{th}$ root of
the identity matrix, up to a global phase factor. To obtain the $T_{k}$'s, one
thus computes the characteristic polynomial $P(\lambda)\equiv\sum_{j=0}%
^{N}a_{j}\left(  \{T_{k}\}_{k=1...N}\right)  \lambda^{j}$ of the matrix
product
\[
e^{-\frac{i}{\hbar}\widehat{H}_{b}T_{N}}\cdot e^{-\frac{i}{\hbar}\widehat
{H}_{a}T_{N-1}}\ldots e^{-\frac{i}{\hbar}\widehat{H}_{a}T_{1}},
\]
and minimizes the function $F_{N}=\sum_{j=0}^{N}\left\vert a_{j}\left(
\{T_{k}\}_{k=1...N}\right)  \right\vert ^{2}$ to $2$ with respect to the
$T_{k}$'s. This minimization turns to be quite easy, due to the fact that a
generic unitary matrix is very close to an $N^{th}$ root of the identity.\ In
fact, numerical work shows that in about 30\% cases of randomly chosen timings
$\left\{  T_{k}\right\}  $ the standard steepest descent algorithm immediatly
finds the global minimum $F_{N}=2$. This fact has roots in the Random Matrix
Theory. Indeed, according to Dyson's law, the eigenvalues of random unitary
matrices tend to repel each other, and are thus very likely to be almost
regularly distributed on the unit circle, as those of an $N^{th}$ root of the
identity, as shown in Fig.\ref{Fig2}. In other words, in the space of
$N\times N$ unitary matrices, the $\widehat{I}^{\frac{1}{N}}$ matrices are present
in abundance, and can be reached from randomly chosen point by small variation
of the timings.

Finally, we define the time vector $\overrightarrow{\tau}^{(0)}$ corresponding
to the identity matrix by simple repetition of $\left\{  T_{k}\right\}  $
\begin{equation}
\tau_{i+\left(  j-1\right)  N}^{(0)}=T_{i}\text{ \ for \ }i,j=1,\ldots,N,
\label{temps}%
\end{equation}
and checks that indeed
\begin{align*}
\widehat{U}\left(  \overrightarrow{\tau}^{(0)}\right)   &  =\underset{N\text{
times}}{\underbrace{\underset{\widehat{I}^{\frac{1}{N}}}{\underbrace{e^{-\frac
{i}{\hbar}\widehat{H}_{b}T_{N}}\cdot e^{-\frac{i}{\hbar}\widehat{H}_{a}T_{N}%
}\ldots e^{-\frac{i}{\hbar}\widehat{H}_{a}T_{N}}}}\ldots\underset{\widehat
{I}^{\frac{1}{N}}}{\underbrace{e^{-\frac{i}{\hbar}\widehat{H}_{b}T_{N}}\cdot
e^{-\frac{i}{\hbar}\widehat{H}_{a}T_{N}}\ldots e^{-\frac{i}{\hbar}\widehat
{H}_{a}T_{N}}}}}}\\
&  =\widehat{I}%
\end{align*}
up to an unimportant global phase factor.

We now have to iteratively determine the time vector $\overrightarrow{\tau}$
from $\overrightarrow{\tau}^{(0)}$. Let us first consider the case of a target evolution close to the identity : in that case, $\widehat{U}_{arbitrary}$ can be written under the form 
\begin{equation}
\widehat{U}_{arbitrary} = \widehat{U}_{\epsilon}\equiv\exp(-i\widehat{\mathcal{H}}\epsilon),
\end{equation}
where $\widehat{\mathcal{H}}$ is an $8\times8$ bounded ($||\widehat{\mathcal{H}}||\leq1$) dimensionless Hermitian Hamiltonian, and $\epsilon>0$ a small parameter. We then look for the variations $\delta \tau_{k}$, determined to
first order in $\epsilon$ by the linear equations
\begin{equation}
\sum_{k=1}^{N^{2}}\frac{\partial\widehat{U}}{\partial \tau_{k}} \left( \overrightarrow{\tau}^{(0)} \right)\,\delta
\tau_{k}=-i\widehat{\mathcal{H}}\epsilon. \label{eqlin}
\end{equation}
Once $\delta\overrightarrow{\tau}$ has been calculated through standard
techniques of linear algebra, we replace $\overrightarrow{\tau}^{(0)}$\ by
$\overrightarrow{\tau}^{(0)}+\delta\overrightarrow{\tau}$ and repeat the same
operation until we obtain $\overrightarrow{\tau}$ which checks $\widehat
{U}\left(  \overrightarrow{\tau}\right)  =\widehat{U}_{arbitrary}$ at the
desired accuracy.

If the evolution $\widehat{U}_{arbitrary}=\widehat{U}_{\epsilon}$ is not close to the identity, that is if $\epsilon$ is not small, one has to divide the work into elementary paths on which the previous method
converges. To this end, we consider an integer $n\geq2$ such that $\left(  \widehat
{U}_{arbitrary}\right)  ^{\frac{1}{n}} = \widehat{U}_{\frac{\epsilon}{n}}$ is attainable from $\widehat{I}$ through our iterative algorithm, and determine in this way the associated time vector
$\overrightarrow{\tau}_{\left(  \frac{1}{n}\right)  }$\ which checks
\[
\widehat{U}\left(  \overrightarrow{\tau}_{\left(  \frac{1}{n}\right)
}\right)  =\widehat{U}_{\frac{\epsilon}{n}}.
\]
Taking $\left(  \widehat{U}_{arbitrary}\right)  ^{\frac{1}{n-1}} = \widehat{U}_{\frac{\epsilon}{n-1}}$ as our new
target, we repeat the same algorithm to compute $\overrightarrow{\tau}_{\left(
\frac{1}{n-1}\right)  }$ such that
\[
\widehat{U}\left(  \overrightarrow{\tau}_{\left(  \frac{1}{n-1}\right)
}\right)  =\widehat{U}_{\frac{\epsilon}{n-1}},
\]
and so on. We progress in this way as long as our algorithm converges : in
general, it stops at a value $n^{\ast}\geq1$, for which the system
Eq(\ref{eqlin}) has no solution. Then, we keep the time vector
$\overrightarrow{\tau}_{\left(  \frac{1}{n^{\ast}}\right)  }$ and simply
repeat the same control sequence $n^{\ast}$ times to achieve the desired
evolution
\[
\underset{n^{\ast}\text{ times}}{\underbrace{\widehat{U}\left(
\overrightarrow{\tau}_{\left(  \frac{1}{n^{\ast}}\right)  }\right)
\cdot\widehat{U}\left(  \overrightarrow{\tau}_{\left(  \frac{1}{n^{\ast}}\right)  }\right)  \ldots\widehat{U}\left(  \overrightarrow{\tau}_{\left(
\frac{1}{n^{\ast}}\right)  }\right)  }}=  \left( \widehat{U}_{\frac{\epsilon}{n^{\ast}}} \right)^{n^{\ast}} = \left[ \left( \widehat{U}_{arbitrary}\right)^{\frac{1}{n^{\ast}}} \right] ^{n^{\ast}}=\widehat{U}_{arbitrary}.
\]

To conclude this section, we provide an equivalent form of our method. Indeed, in the
previous paragraphs we fixed the amplitudes of the perturbations once for all and considered the pulse timings $\tau_{k}$ as our free control parameters.
But we also might have chosen to apply $N^{2}$ pulses of same duration
$\tau=\frac{T}{N^{2}}$, where $T$ is the total control sequence duration, and
taken the amplitudes as our free control variables. In other words, we might
have applied the following Hamiltonian
\begin{equation}
\widehat{H}(t)=\left\{
\begin{array}
[c]{c}%
\widehat{H}_{2k+1}=\widehat{H}_{0}+C_{2k+1}\widehat{P}_{a}\\
\widehat{H}_{2k}=\widehat{H}_{0}+C_{2k}\widehat{P}_{b}
\end{array}
\right.
\end{equation}
where the amplitudes $\left\{  C_{k}\right\}  _{k=1,\ldots,N%
{{}^2}%
}$ play the role of adjustable control parameters. The evolution operator would have thus
taken the form
\begin{equation}
\widehat{U}\left(  \left\{  C_{k}\right\}  \right)  =e^{-\frac{i}{\hbar
}\left(  \widehat{H}_{0}+C_{N%
{{}^2}%
}\widehat{P}_{b}\right)  \tau}\cdot e^{-\frac{i}{\hbar}\left(  \widehat{H}%
_{0}+C_{N%
{{}^2}%
-1}\widehat{P}_{a}\right)  \tau}\ldots e^{-\frac{i}{\hbar}\left(  \widehat
{H}_{0}+C_{1}\widehat{P}_{a}\right)  \tau},
\end{equation}
and our problem of control would have boiled down to finding the vector $\overrightarrow
{C}=\left(
\begin{array}
[c]{c}%
C_{1}\\
\vdots\\
C_{N^{2}-1}\\
C_{N^{2}}%
\end{array}
\right)  $ such that $\widehat{U}\left(  \overrightarrow{C}\right)
=\widehat{U}_{arbitrary}$.

The method remains the same as previously. First, one determines the
parameters $\left\{  c_{k}\right\}  _{k=1,\ldots,N}$ such that
\[
e^{-\frac{i}{\hbar}\left(  \widehat{H}_{0}+c_{N}\widehat{P}_{b}\right)  \tau
}\cdot e^{-\frac{i}{\hbar}\left(  \widehat{H}_{0}+c_{N-1}\widehat{P}%
_{a}\right)  \tau}\ldots e^{-\frac{i}{\hbar}\left(  \widehat{H}_{0}%
+c_{1}\widehat{P}_{a}\right)  \tau}=\widehat{I}^{\frac{1}{N}},
\]
by minimizing the functional $\sum_{j=0}^{N}\left|  a_{j}\left(
\{c_{k}\}_{k=1\ldots N}\right)  \right|  ^{2}$ to $2$, where $\left\{  a_{j}\right\}
$ denote the coefficients of the characteristic polynomial of the matrix
product
\[
e^{-\frac{i}{\hbar}\left(  \widehat{H}_{0}+c_{N}\widehat{P}_{b}\right)  \tau
}.e^{-\frac{i}{\hbar}\left(  \widehat{H}_{0}+c_{N-1}\widehat{P}_{a}\right)
\tau}\ldots e^{-\frac{i}{\hbar}\left(  \widehat{H}_{0}+c_{1}\widehat{P}%
_{a}\right)  \tau},
\]
and we set
\begin{equation}
C_{i+\left(  j-1\right)  N}^{(0)}=c_{i}\text{ \ for \ }i,j=1,\ldots,N,
\end{equation}
such that $\widehat{U}\left(  \overrightarrow{C}^{\left(  0\right)
}\right)  =\widehat{I}$. 

Then, one iteratively approaches the desired vector
$\overrightarrow{C}$ in the same way as previously. For a target evolution $\widehat{U}_{arbitrary} = \widehat{U}_{\epsilon}=\exp(-i\widehat{\mathcal{H}}\epsilon)$ close to
the identity (\emph{i.e.} for $\epsilon$ small), one computes the variations $\delta C_{k}$ to
first order in $\epsilon$ by solving the linear equations
\begin{equation}
\sum_{k=1}^{N^{2}}\frac{\partial\widehat{U}}{\partial C_{k}}\left( \overrightarrow{C}^{(0)} \right)\,\delta
C_{k}=-i\widehat{\mathcal{H}}\epsilon.
\end{equation} 
Then we replace $\overrightarrow{C}^{(0)}$ by $\overrightarrow{C}^{(0)}+ \delta \overrightarrow{C}$ and repeat the same operation, and so on, until we get $\overrightarrow{C}$ which checks $\widehat{U}\left(  \overrightarrow{C}\right)=\widehat{U}_{arbitrary}$ with the desired accuracy. 

For a target evolution $\widehat{U}_{arbitrary} = \widehat{U}_{\epsilon}$ far from the identity (\emph{i.e.} for $\epsilon$ finite), one considers an arbitrary integer $n \geq 1$ such that $\left(\widehat{U}_{arbitrary}\right)
^{\frac{1}{n}} = \widehat{U}_{\frac{\epsilon}{n}}$ can be reached from the identity through the previous iterative
algorithm, and computes $\overrightarrow{C}^{(\frac{1}{n})}$ which checks
\[
\widehat{U}\left(  \overrightarrow{C}^{\left(  \frac{1}{n}\right)  }\right)
=\widehat{U}_{\frac{\epsilon}{n}}.
\]
Then, one calculates in the same way the vector $\overrightarrow{C}^{\left(
\frac{1}{n-1}\right)  }$ such that
\[
\widehat{U}\left(  \overrightarrow{C}^{\left(  \frac{1}{n-1}\right)  }\right)
=\widehat{U}_{\frac{\epsilon}{n-1}},
\]
and so on, until one obtains the limiting value $n^{\ast}\geq1$ beyond which the algorithm
fails to converge. Finally, one gets the desired evolution by repeating the same control sequence $n^{\ast}$ times, the amplitudes of which are given by the vector $\overrightarrow{C}^{\left( \frac{1}{n^{\ast}} \right)}$.

\section{Conclusion}

In this paper, we presented a universal method for controlling the evolution operator of any quantum system. This control scheme, called the non-holonomic control, consists in alternately applying two physical perturbations which check the bracket generation condition. The role of tunable control parameters may be played either by the durations of the interaction pulses or by the strengths of the perturbations during these pulses. The calculation of the control parameters is achieved through an algorithm, that we presented here, the convergence of which is assured by remarkable properties of the $N^{th}$ roots of the identity matrix. 

In the following paper, we show how the non-holonomic control technique can be used in the context of quantum computation in order to build controlled quantum devices.

\end{document}